\documentclass{IOS-Book-Article}

\usepackage{mathptmx}
\usepackage{soul}\setuldepth{article}
\def\hb{\hbox to 11.5 cm{}}
\usepackage{amsmath}
\usepackage{scrextend}
\usepackage{graphicx}
\usepackage{comment}

\begin{document}

\pagestyle{headings}
\def\thepage{}
\begin{frontmatter}              % The preamble begins here.

%\pretitle{Pretitle}
\title{A prototype hybrid prediction market for estimating replicability of published work}

%\markboth{}{April 2023\hb}
%\subtitle{Subtitle}

\author[A]{\fnms{Tatiana} \snm{Chakravorti}}, 
\author[A]{\fnms{Robert} \snm{Fraleigh}},
\author[A]{\fnms{Timothy} \snm{Fritton}},
\author[A]{\fnms{Michael} \snm{McLaughlin}},
\author[A]{\fnms{Vaibhav} \snm{Singh}},
\author[A]{\fnms{Christopher} \snm{Griffin}},
\author[A]{\fnms{Anthony} \snm{Kwasnica}},
\author[B]{\fnms{David} \snm{Pennock}},
\author[A]{\fnms{C. Lee} \snm{Giles}},
\author[A]{\fnms{Sarah} \snm{Rajtmajer}\thanks{Corresponding Author: smr48@psu.edu}}

%\runningauthor{T. Chakravorti et al.}

\address[A]{The Pennsylvania State University}
\address[B]{Rutgers University}

\begin{abstract}
%Despite its successes, Artificial intelligence (AI) still suffers from important limitations, particularly for complex tasks where common sense, creativity, and intuition are required, or when learning from limited data. The promises and challenges of AI have motivated work exploring frameworks for human-AI collaboration which may bring together human intuition and machine rationality to tackle today’s grand challenges effectively.
We present a prototype hybrid prediction market and demonstrate the avenue it represents for meaningful human-AI collaboration.  We build on prior work proposing artificial prediction markets as a novel machine learning algorithm. In an artificial prediction market, trained AI agents (bot traders) buy and sell outcomes of future events. %In the work we will overview here, these outcomes are the results of replication studies of published research. 
Classification decisions can be framed as outcomes of future events, and accordingly, the price of an asset corresponding to a given classification outcome can be taken as a proxy for the system’s confidence in that decision. 
%The most exciting promise of artificial markets, we suggest, is the novel avenue they provide for human-AI collaboration. 
By embedding human participants in these markets alongside bot traders, we can bring together insights from both. In this paper, we detail pilot studies with prototype hybrid markets for prediction of replication study outcomes. We highlight challenges and opportunities, share insights from semi-structured interviews with hybrid market participants, and outline a vision for ongoing and future work.  
\end{abstract}

\end{frontmatter}
%\markboth{April 2023\hb}{April 2023\hb}
%\thispagestyle{empty}
%\pagestyle{empty}

\section{Introduction}

A nascent literature is exploring artificial prediction markets -- numerically simulated markets, populated by artificial agents (bot traders) for supervised learning of probability estimators \cite{barbu2012introduction}. Early work has demonstrated the plausibility of using a trained market as a supervised learning algorithm, achieving comparable performance to standard approaches on simple classification tasks \cite{barbu2012introduction,barbu2013artificial,jahedpari2014artificial,nakshatri2021design}.  We suggest the most promising opportunity afforded by artificial prediction markets is eventual human-AI collaboration -- a market framework that supports human traders participating alongside agents to evaluate outcomes. In an initial study \cite{chakravorti2022}, we have outlined the theoretical foundation for such a \emph{hybrid prediction market} and simulated simple human-like behaviors in this setting.%, laying groundwork for work in progress embedding real human participants into a synthetic prediction market. 

The hybrid markets we describe here aim to predict the outcomes of replication studies in the social and behavioral sciences. The study of reproducibility, replicability, and robustness of published scientific findings has gained widespread attention in the social sciences and beyond. A number of large-scale replication projects \cite{errington2014science,open2015estimating,camerer2016evaluating,camerer2018evaluating,klein2014investigating,klein2018many,cova2021estimating} have reported successful replication rates anywhere between 36\% and 78\% and have sparked high-profile debate about the reliability of published findings \cite{baker20161,gilbert2016comment,fanelli2018opinion,redish2018opinion}. The task of forecasting replication outcomes appears to be an ideal candidate for human-AI collaboration. Both human and machine-centered efforts have shown promise but neither has achieved good performance, e.g., \cite{dreber2015using,yang2020estimating}. Indeed, replication prediction appears to be a problem for which the scale and scope of machine-driven approaches is necessary but capturing the intangible wisdom of experts in the field remains elusive.

Our prior work \cite{rajtmajer2022synthetic} has developed and deployed synthetic prediction markets for the task of replication prediction. We further this work here to explore a hybrid market scenario. Our work in progress is scaffolded by three research questions, the answers to which will inform larger-scale development of human-AI hybrid prediction markets as a novel avenue for creative peer review.  
%Work in progress detailed below is the next step in a line of work which has, thus far, developed and deployed synthetic prediction markets for the task of replication prediction \cite{rajtmajer2022synthetic}. In the hybrid scenario, we explore three research questions, the answers to which will inform larger-scale development of human-AI hybrid prediction markets as a novel avenue for creative peer review.

\begin{addmargin}[1em]{2em}
\textbf{RQ1}: How does human participation in a synthetic prediction market impact market performance vs. the purely synthetic setting?
\end{addmargin}

\begin{addmargin}[1em]{2em}
\textbf{RQ2}: In the context of replication prediction, what features matter most to human participants? How do participants formulate trading strategies? % features are important to decide Research Reproducibility? How did the participants formulate the trading strategy? What are the challenges faced during the hybrid market test run?
\end{addmargin}

\begin{addmargin}[1em]{2em}
\textbf{RQ3}: What outstanding challenges need to be addressed prior to large-scale deployment of hybrid prediction markets for replication prediction? 
\end{addmargin}

\noindent Overarchingly, these questions serve the broader aim of understanding how hybrid human-AI technologies can help us evaluate reproducibility, replicability, and robustness of published scientific findings.  %Need for action when it comes to research reproducibility. How technology can support research reproducibility in today's academic practices? What rewards could be conductive to reproducible research practices?
Following, we discuss insights from beta testing a hybrid market for replication prediction. We also discuss participants' perspectives based on follow-up interviews. We close with a vision for further work in this area.

\section{Data}
%Training data and test data .... XXX \footnote{Training data is available at \url{https://github.com/Tatianachakravorti}.}
%Our work builds upon a dataset of XXX known replication study outcomes from the Replication Project Psychology \cite{}, the Social Science Replication Project (SSRP) \cite{}, XXX... 
Algorithmic agents were trained on outcomes of $400$ replication studies and expert evaluations of published findings in the social and behavioral sciences, spearheaded by the Center for Open Science and the University of Melbourne for DARPA's Systematizing Confidence in Open Research and Evidence (SCORE)\footnote{See \url{https://www.darpa.mil/program/systematizing-confidence-in-open-research-and-evidence}.} program. Twelve additional replication study outcomes served as test data for the hybrid market events.\footnote{Coordinated release of all data from the SCORE program is planned in late 2023. The subset of data used in our analyses will be linked at \url{https://github.com/Tatianachakravorti} as soon as available.} Domains and journals from which findings were selected, as well as procedures for replications and expert evaluations are detailed in \cite{alipourfard2021systematizing}.

%It was critical to experimental integrity that replication outcomes used to test the hybrid markets were not already published so that participants were not able to access ground truth (data leakage) during the experiment. As such, we were given a set of 12 replication outcomes that have not yet been published from the Center for Open Science. These outcomes were a small subset of massive-scale DARPA SCORE effort \cite{alipourfard2021systematizing}.

%\subsection{Features From Research Articles}

%Our artificial prediction market is trained using a set of features extracted from a dataset of 300 Social Behavioral Science scholarly papers and their metadata, each labeled as replicable or not-replicable. The feature set spans five categories and includes semantic, bibliometric, author-related, venue-related, and statistical information.

%In this research, we have used the extracted feature set as our dataset to train the artificial prediction market. The feature extraction describes five categories of features extracted from the 300 Social Behavioural Science scholarly papers and their metadata, each labeled as replicable or not replicable. They are semantic, bibliometric, author-related, venue-related, and statistical information. 

Full text of train and test papers were passed through a feature extraction pipeline to obtain semantic, bibliometric, and statistical information. Specifically, 41 features, e.g., reported p-values, author names, author count, venue, acknowledgment of funding, were extracted for each research claim in question. See \cite{wu2021predicting} for further detail.

%\subsection{Participant Interviews}

%For the Hybrid Market pilot study, we recruited participants from three research backgrounds: Business, Information Science, and Psychology. Of the 33 participants originally recruited for the hybrid market pilot study, 8 participants continued to an interview study. These participants all possessed a hands-on experience with market performance. The main inclusion criteria for this study, were researchers and professors in the social and behavioral sciences as defined in the scope of our project (psychology, sociology, political science, education, economics, business, political science, anthropology, communications) either hold a Ph.D. in their discipline or are currently enrolled in a Ph.D. program. We also considered engineering background students as well in this pilot study who are from Information Science and Technology. All the candidates should be of age 18 years and older. 

\section{Methodology}

\subsection{Artificial Prediction Market Model}

Given the resources required to run high-powered replication studies, researchers have sought other approaches to assess confidence in published claims and have looked to creative assembly of expert judgement as one opportunity. Initial evidence has supported the promise of prediction markets in this context. Simple, binary option prediction markets have outperformed survey-based approaches in predicting outcomes for a number of high-profile replication projects \cite{dreber2015using,camerer2016evaluating,camerer2018evaluating,forsell2019predicting,gordon2020replication,gordon2021predicting}. In these markets, assets corresponding to future events (i.e., results of replication studies) can be bought and sold thereby manipulating underlying asset prices. These asset prices can be interpreted as probabilities \cite{manski2006interpreting,wolfers2006interpreting} thereby providing a mechanism for event forecasting. 

\begin{figure}
    \vspace{-0.6cm}
    \centering
    \includegraphics[width=\textwidth]{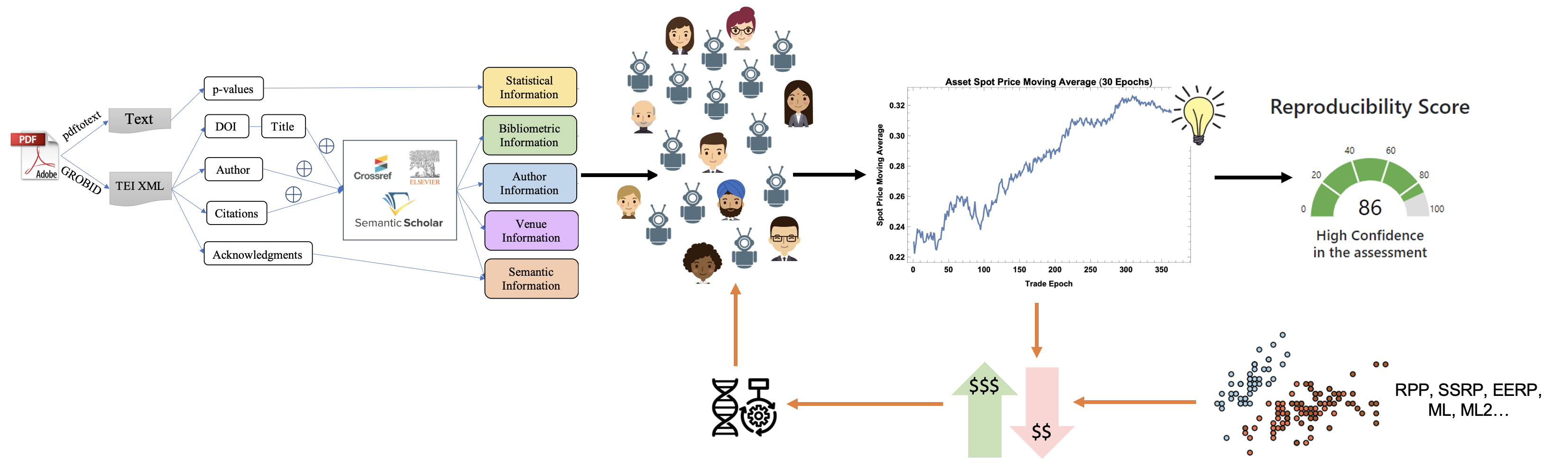}
    \caption{\textbf{Schematic representation of market for replication prediction.} Step 1: Features are extracted from full-text of a paper of interest (see \cite{wu2021predicting}). Step 2: Extracted features are passed to algorithmic traders. Step 3: Algorithmic traders buy and sell contracts representing `will replicate' or 'will not replicate' outcomes of a replication study associated with the primary claim of the paper of interest. \emph{Note: in the hybrid market scenario, human traders participate alongside bots.} Trading manipulates the underlying asset prices via logarithmic market scoring rule (see \cite{nakshatri2021design}). (\emph{training phase, orange arrows}) Step 3. At market close, the outcome of the replication study is revealed. Algorithmic traders profit or lose money based on the total value of the assets they hold. Traders who profit are allowed to reproduce, mutate, and remain in the market via genetic algorithms. (\emph{test phase, black arrows}) Step 3. The price of a `will replicate' asset at the time of market close is given as a proxy for the market's prediction.}
    \label{fig:schematic}
    %\vspace{-0.cm}
\end{figure}

The base model of this work is a simple artificial prediction market populated by algorithmic agents (bot traders) whose decisions to buy contracts are based on extracted features from full text of published work and associated metadata and who are trained on ground-truth replication outcomes using a genetic algorithmic approach. Further detail on the mathematical formulation of the artificial market, bot training procedures, and feature extraction from scholarly manuscripts are detailed in prior work \cite{nakshatri2021design,wu2021predicting,rajtmajer2022synthetic}. A schematic representation of the market and training processes is given in Figure \ref{fig:schematic}.

\subsection{Hybrid Market Experimental Design}

\begin{figure}
    \vspace{-0.9cm}
    \centering
    \includegraphics[width=\textwidth]{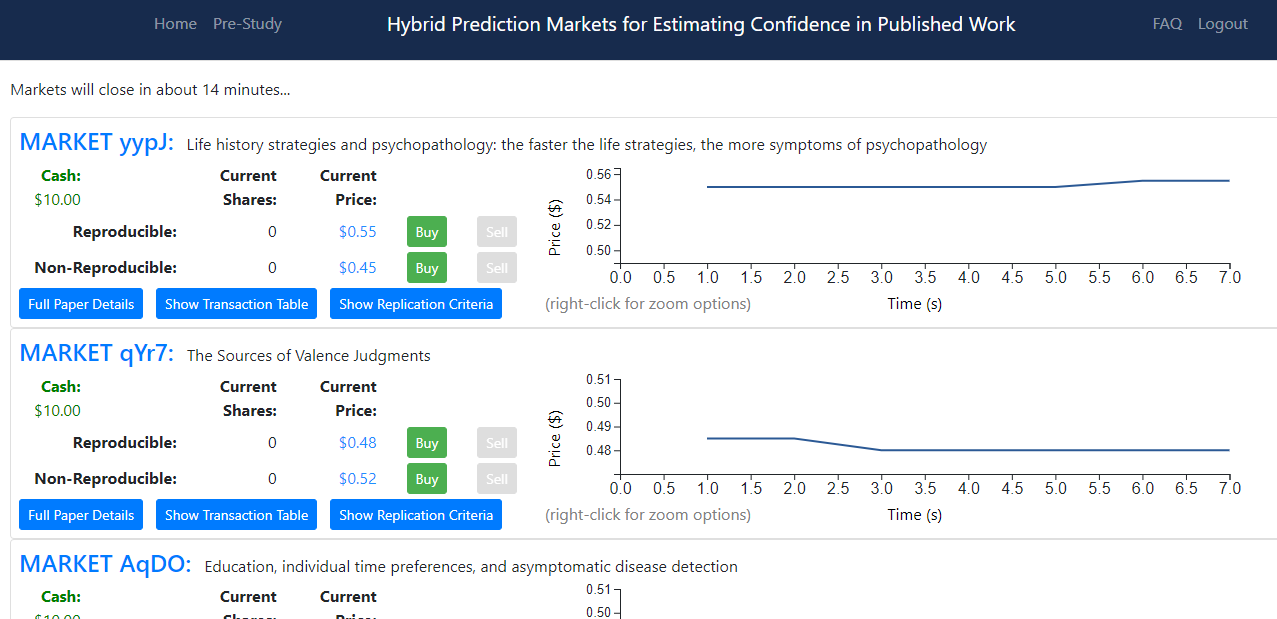}
    \caption{\textbf{Interactive web application.} Human participants are provided with information about the replication study and full text of the associated paper. We provide them with initial cash to invest. They may buy and sell contracts representing `will replicate' and `will not replicate' outcomes of the study via the app.}
    \label{fig:UI}
    \vspace{-0.3cm}
\end{figure}

In ongoing research, we have built a beta-tested a platform to enable bot interactions with real human participants, i.e., a hybrid market scenario, for replication prediction. The platform includes a web server that hosts a pre-trained artificial market and several API endpoints that: 1) enable artificial agents to buy assets; 2) enable human participants to buy and sell assets; 3) manage transaction bookkeeping and experiment statistics. The platform also includes an interactive web application for human participants to buy and sell assets intuitively (see Figure \ref{fig:UI}).

While the theoretical foundation of the artificial market is framed continuously, the deployed artificial and hybrid markets are discrete, iterating every one second. Market transactions were managed using a queuing system, first evaluating all participating agents and then any human transactions using a first-in-first-out rule. Both artificial agents and human participants were limited to single-share transactions. Upon each market transaction, the market price for all assets was updated using a logarithmic market scoring rule (LMSR) \cite{nakshatri2021design}. During each market iteration, a stochastic sampling of available artificial agents were selected to participate; the sampling rate was reduced to ~5\% of the artificial agent population to align market price convergence with a 2-hour timeline (7200 market iterations) for the hybrid market scenario. 

%We recruited participants from three research backgrounds: Business school, Information Science, and Psychology department. Of the 33 participants originally recruited for the hybrid market pilot study, 8 participants continued to an interview study. These participants all possessed a hands-on experience with market performance. The main inclusion criteria for this study, were researchers and professors in the social and behavioral sciences as defined in the scope of our project (psychology, sociology, political science, education, economics, business, political science, anthropology, communications) either hold a Ph.D. in their discipline or are currently enrolled in a Ph.D. program. We also considered engineering background students as well in this pilot study who are from Information Science and Technology. All the candidates should be of age 18 years and older. 

%We ran four markets simultaneously over a 2 hour period, with one market providing a payout for those participants who held stock in the correct asset at the end of the experiment. Artificial agents were initialized with a large sum of cash (\$500) so that agent behavior, in aggregate, would mimic real-world market forces. Human participants received a smaller sum, \$125, segmented across each market (\$25 per market) to spend. The market had two assets: reproducible and non-reproducible. Artificial agents were restricted to only buy options for both asset types; human participants were allowed to buy and sell both asset types.

Hybrid markets were beta-tested in three separate events. Instiutional Review Board (IRB) approval was obtained prior to these events. Each of the three events involved four markets; that is, in each event, participants were given the opportunity to evaluate and buy/sell outcomes of four distinct replication studies. We recruited participants for each, respectively, with three research backgrounds: graduate students at Penn State's Smeal School of Business (Event 1: April 2022, 9 participants), graduate students at Penn State's College of Information Science and Technology (Event 2: June 2022, 10 participants), and graduate students from the Department of Psychology at Penn State (Event 3: July 2022, 14 participants). %Each participant  who either hold a Ph.D. in their discipline or are currently enrolled in a Ph.D. program. All the candidates were older than 18 years. 
 %We recruited subjects for the study over LinkedIn, and social media (e.g. Twitter) and through direct email engagement with on-campus departments and also outside the campus. We recruited 14 participants for psychology, 10 from IST, and 9 participants from business school for these three sessions to test the Hybrid market performance. 
 Event 1 was conducted in-person at the Laboratory for Economics, Management \& Auctions on campus and markets were open for one hour. Events 2 and 3 were held virtually and markets were open for two hours. Participants in all events were given $\$20$ for their time and an additional $\$25$ to invest in the markets.  % remotely over a two hour period; %The session with the business school participants was conducted in-person at the Laboratory for Economics, Management \& Auctions on campus and lasted one hour. In all sessions, one market was randomly selected for payment. 
 At the conclusion of each event, one market was randomly selected for payment. The selected market along with the replication outcome for that market was revealed to participants. %that determined asset valuation. 
 Assuming a minimal activity requirement was satisfied, each participant was paid the value of their asset holdings in the selected market and any remaining (uninvested) cash. The average participant earning was \$42.62. 

%The market iterated state every second and transactions 

%To receive a payout for the selected market, we required a minimum of three transactions in that market. Payout amounts were \$1 for a correctly chosen asset, \$0 for an incorrectly chosen asset. The ground-truth for a publication’s reproducibility was reported by the Open Science Foundation.

%Our hybrid market study was submitted to and received approval from our institutional ethics and IRB team. We advertised the study over LinkedIn, social media (e.g. Twitter) and through direct email engagement with on-campus departments and also off campus. We recruited thirty-three participants (14 psychology, 10 information science, and 9 business) for these three experiment sessions to test the Hybrid market performance.

\subsection{Participant Surveys}
Hybrid market participants were asked to complete pre- and post-experimental surveys. The pre-market survey assessed participants' research background and familiarity with scholarly work on reproducibility and replicability. In addition, the pre-market survey asked participants to  provide feedback on each paper they were about to evaluate during the hybrid market event. %and specific questions related to the publication reproducibility (likelihood, confidence, final price) for each experimental market. 
The post-market survey asked participants to describe their trading strategy, specific features of the studies, e.g., sample size, author reputation, that guided their predictions. We asked whether they were surprised by the market outcome, and whether they had changed their mind from their original assessments.\footnote{Survey instruments are shared at \url{https://github.com/Tatianachakravorti}.}

\subsection{Participant Interviews}

Table~\ref{tab:datasummary} provides a summary of hybrid market outcomes, alongside corresponding artificial market outcomes for each test paper. For each market, the final price (in dollars) is reported for an asset that pays \$1 if the finding was successfully replicated and \$0 if not, i.e., the `will replicate' asset.\footnote{The market price of the `will not replicate' asset is always 1 minus the price of this asset by construction.} Thus, we classify a market prediction as `correct' if the price is greater (less) than .5 and the paper was (not) reproducible. Absolute error (AE) is the absolute difference between the final price and the value of the asset (0 or 1). Markets with consistently lower AE are considered better predictors. 
We also conducted 30-minute semi-structured one-on-one interviews via Zoom teleconferencing with $8$ hybrid market participants who responded to our request for additional feedback. Interviewees were active researchers with a PhD or currently enrolled in a PhD program.\footnote{Six of the 8 interviewees were participants in hybrid market experiments run in October 2022. They were not part of the three market events reported in Table \ref{tab:datasummary}. October events followed the same 2-hour format and used the same platform as the market events in April, June and July 2022.} %We emailed 15 participants and those who responded for further discussion were selected for this study. All participants were interviewed individually about their experiences. Building on our Hybrid Market test run, 
Our interview protocol included questions about their experience in the market: features of papers and studies that helped them to estimate replicability; any strategic behavior while trading; general impressions of the platform. We also asked them about perceptions of reproducibility and replicability in their field, %. We asked the extent to which their peers were aware of challenges to reproducibility in their field, 
how technology might support reproducibility, and incentives for replication. %; general questions about their experience during the hybrid market test run. 
Participants received $\$20$. %The range of interviewee perspectives are included in the results section.

\section{Results}

%This section describes the results from the hybrid pilot study and also from the semi-structured interviews. 

\subsection{RQ1: Hybrid market performance}

Table~\ref{tab:datasummary} provides a summary of hybrid market outcomes, alongside corresponding artificial market outcomes for each test paper. For each market, the final price (in dollars) is reported for an asset that pays \$1 if the finding was successfully replicated and \$0 if not, i.e., the `will replicate' asset. \footnote{The market price of the `will not replicate' asset is always 1 minus the price of this asset by construction.} Thus, we classify a market prediction as `correct' if the price is greater (less) than .5 and the paper was (not) reproducible. Absolute error (AE) is the absolute difference between final price and value of the asset (0 or 1). Markets with consistently lower AE are considered better predictors. 

Hybrid market predictions were globally more accurate than predictions of the artificial markets (mean AE .497 vs .552). In 9 of 12 markets, AE was lower in the hybrid setting. A Wilcoxon signed ranks test fails to reject the null hypothesis that the distributions of errors between the hybrid and artificial markets are the same ($z =-1.373$, $p-value = 1.83$), likely due to the low power of the small sample. In no instance is the hybrid market price incorrect when the synthetic price was correct; in one instance the hybrid market flipped an incorrect prediction of the synthetic market to correct. 

Artificial markets are vulnerable to lack of participation; agents will not participate if they have not seen a sufficiently similar training point (paper). In practice, this may leave some test points unevaluated. We have observed this in prior work \cite{rajtmajer2022synthetic} and that is the case here in five of twelve markets.\footnote{In the absence of trading, the market price for both assets is .5.} In the hybrid setting, human participation can support a prediction. And notably, we observed that \emph{the presence of human traders often induced greater participation amongst algorithmic agents.} In four of the five markets inactive in the artificial setting, the presence of human traders induced agent trades; only in hybrid market E2M2 were agents completely inactive. %Further, {\color{magenta}the volume of agent-based trades increased from an average of X per market in the fully artificial setting to Y per market in the hybrid setting}. 

These small sample results indicate that small numbers of informed/expert human traders do not have an obviously deleterious effect on market performance and might even improve accuracy. Likewise, the trading activity of human traders has the potential benefit of triggering trading  amongst algorithmic agents.

\begin{table*}[t]
  \caption{\textbf{Experimental data summary.} R/NR: replicated/not replicated; C/NC: correct/not correct;  price: final price; pred: prediction; AE: absolute error; and, -- denotes no agent participation. }
  \label{tab:datasummary}
  \begin{tabular}{c|c|ccc|ccc}
  %\toprule
 &  & Hybrid & Hybrid & Hybrid & Artif & Artif & Artif\\ 
  Market  & Outcome & price & pred & AE & price & pred & AE\\ \hline
    %\midrule
Event 1 Market 1 (E1M1) &    R & 0.66 &    C & 0.34 & 0.41 &       NC & 0.59\\
Event 1 Market 2 (E1M2) &    R & 0.36 &    NC & 0.64 & 0.5 &   -- & 0.5\\
Event 1 Market 3 (E1M3) &    R & 0.64 &    C & 0.36 & 0.52 &       C & 0.48\\
Event 1 Market 4 (E1M4) &    NR & 0.72 &    NC & 0.72 & 0.5 &       -- & 0.5\\ \hline
Event 2 Market 1 (E2M1) &    R & 0.38 &    NC & 0.62 & 0.41 &       NC & 0.59\\
Event 2 Market 2 (E2M2) &  R & 0.58 &    C & 0.42 & 0.5 &       -- & 0.5\\
Event 2 Market 3 (E2M3) &    R & 0.8 &    C & 0.2 & 0.52 &       C & 0.48\\
Event 2 Market 4 (E2M4) &    NR & 0.47 &    C & 0.47 & 0.5 &       -- & 0.5\\ \hline
Event 3 Market 1 (E3M1) &    R & 0.61 &    C & 0.39 & 0.5 &  -- & 0.5\\
Event 3 Market 2 (E3M2) &    R & 0.47 &    NC & 0.53 & 0.46 & NC & 0.54\\
Event 3 Market 3 (E3M3) &    NR & 0.76 &    NC & 0.76 & 0.86 & NC & 0.86\\
Event 3 Market 4 (E3M4) &    R & 0.49 &    NC & 0.51 & 0.42 &       NC & 0.58\\ \hline
%\bottomrule
\end{tabular}
\end{table*}

\subsection{RQ2: Human participants' evaluative criteria and trading strategies}

Findings in support of RQ2 primarily derive from participant interviews.  
%This section presents the results generated from the interview studies as qualitative data. 
We analyzed interview transcripts %of the interview audio files using an inductive analysis 
using an inductive approach guided by specific evaluation objectives. To analyze and code these transcripts we have used the Taguette software \cite{rampin2021taguette}. %We have gone through the transcripts line by line and labeled the ideas. This thematic analysis uncovered few major patterns or themes for this study. %Also, we have tried to find out the similarities between the participants' thinking and merged them together under same findings. We have focused on the prominent findings throughout the interview transcripts which are significant for our research questions. At the end, the similar codes are grouped together to find a higher level of understanding or categories. Many times we have revisited the interview transcripts if there are any doubts. The following subsections present each of these themes and provide sample quotes as examples/proofs. 
Following, we provide key themes and exemplar quotes.

%\subsubsection{Important features for Research Reproducibility}
\subsubsection{Study motivation, design, and reported outcomes}

%We wanted to understand the features that influenced participants' decisions about the reproducibility of that paper. This section reflects all the important features raised by the participants and according to our analysis there is a huge similarity in their thinking. Based on their interviews all the 
All eight interviewed participants mentioned methodology, sample size, p-values and soundness of research questions as prominent evaluative criteria.  %reproducibility check. For example, 

\begin{addmargin}[1em]{2em}
\emph{I focused, specifically on the size of the sample, the diversity of the sample, and the complexity of the question, in order to predict how reliable it would be\ldots so if the sample was small if the population wasn't diverse and the question was complex\ldots This seems to have a lower likelihood of reproducing.} [Participant 2]
\end{addmargin}

\vspace{0.2cm}

\begin{addmargin}[1em]{2em}
\emph{First, there's the prior probability. Does it sound like a possible idea or not?\ldots Second is the sample size\ldots P-values are important\ldots I was looking for pre-registrations or open data.} [Participant 5]
\end{addmargin}
%…... and I was also looking for dates. If it was like late two thousand very early, two thousand and ten kinds of pre-replication crisis, I thought. Well, people might not have been paying attention to this yet, and then the last one was looking at the manipulation check.

%"I've checked their methodology part, and the number of samples ... And if they use like reasonable methodology or not. I also checked the P. Values"[Participant 6]

%"First sample size, then the possibility of the design second, and maybe effect sizes. But honestly, I didn't put too much weight on that" [Participant 7]

\subsubsection{Journal and author reputation}

%This section reveals how most of the participants have some effect on their decisions based on the high impact factor journals while reading the papers which is very interesting to observe from this study. 
Six participants noted an impact of journal reputation on their prediction. However, most participants reported that they did not consider the papers' authors; one participant noted an exception for one paper.

\begin{addmargin}[1em]{2em}
\emph{I guess the journal itself.... I know some of the journals have tended to have higher-quality kind of articles. So I’m sure that that affected my thought process to some extent but really the main thing was, you know, sample size, methodology, and theoretical orientation.} [Participant 1].
\end{addmargin}

\vspace{0.2cm}

%\begin{addmargin}[1em]{2em}
%\emph{I did check the journal names from what I could tell, they all seem to be like mid-tier-ish journals in terms of impact score. So I didn't take that as too much of a factor. But it did cross my mind.} [Participant 2].
%\end{addmargin}

%\vspace{0.2cm}

\begin{addmargin}[1em]{2em}
\emph{Yeah, I am not sure about the journals in general but in one case I remember that I forgot now the name of the paper. But since I know who is the author and I believe that he's good.%I thought his work is reproducible. Also then that it's tended to bias me toward the decision.
} [Participant 4].
\end{addmargin}

%"I didn't pay attention to citations. I believe I was biased in favor of high-ranking journals in my experience.... And yeah, I was influenced by that."[Participant 7]

\subsubsection{Trading strategy}

%This section represents the trading strategy of the participants during the hybrid market test run. The overall observation is 
Five participants out of eight reported having a probability in mind for the replicability of each finding prior to the markets opening, and made initial investments based on this probability. % according to their decision towards reproducibility. 
However, several mentioned that they later changed their mind when the market start trending in opposite direction to their initial prediction.

%"In the first market, I was trying to trade on the basis of my own expectations .... and things were kind of moving in a certain direction. And I was wrong about the main article. So I was buying on the basis of what I thought would be the case and waiting for the price to move in a direction that was favorable for the case that I was looking at. but I think, it was not as good of a strategy. And so the second time around. I sort of was listening to the market itself..... I thought the article was replicable, but the market was trending in the direction of not replicable. I tried to buy as many of the not replicable ones as I could"[Participant 1].

\begin{addmargin}[1em]{2em}
\emph{I have a probability in mind\ldots so if I think that it should be seventy-five, and the price is sixty percent then I'm willing to buy because I think I'm making fifteen cents. But if it's trading at forty and I think that the probability of replication is twenty, I'm happy to buy because I think there I'm making twenty cents uh on the dollar or on the share or whatever.} [Participant 5].
\end{addmargin}

\vspace{0.2cm}

\begin{addmargin}[1em]{2em}
\emph{I wrote down prior to the market what would be in my eyes a reasonable estimate of the probability of replicability. Then I observe market prices and try to make trades that would bring the market price closer to my reasonable estimate\ldots But then observing market prices, and in some markets observing trends that, like went from fifty-fifty in the opposite direction. I was starting to reconsider whether I’m wrong with my initial estimate. }% Yeah. Then, after reconsidering, I kind of revised my opinion, and for one market I was so sure that I tried to heavily try it against the market. Because I was so sure that it would actually replicate that I tried to bring it up.} 
[Participant 7]
\end{addmargin}

\subsubsection{Impact of agent participation on decisions}

All participants reported that the presence of agents in the market had no impact on their behavior because they had no information about the agents' training or behavior. %what features the agents were learning or any other information about the training of these agents, so the presence of agents in the system had no impact on their decisions.% due to the agent's participation in the market. 

\subsection{RQ3: Perspectives on human-AI technologies for replication prediction}

\subsubsection{Concerns about reproducibility and replicability}

%The overall understanding from this section is that everyone feels that there is a need for research reproducibility and how it is important but the practices of doing that type of research are still very less. All the participants care about the reproducibility of the outcomes when they look for papers in their research domains. But the new generation researchers are more into this rather than the previous generation. For example,
All participants interviewed reported concern about reproducibility and replicability. Although we expect this is related to self-selection bias for our study. Participants reported variability in awareness around these issues in their field.

\begin{addmargin}[1em]{2em}
\emph{I guess maybe thirty percent of people in my field have pretty strong concerns about it. }[Participant 1]
\end{addmargin}
%"Yes, of course. Yeah, since I know about the reproducibility crisis. And I think that it is very important."[Participant 4]

\vspace{0.2cm}

\begin{addmargin}[1em]{2em}
\emph{People my age tend to be very aware of this. professors a generation or two above me. There are. They're at least aware of it. They may not uh accept that. It's an issue, but they are, I think, generally, people aware of the issue. }[Participant 5]
\end{addmargin}

\vspace{0.2cm}

\noindent Participants acknowledged lack of incentives for practices heralded by the open science movement (e.g., preregistration) and lack of venues for publishing replication studies.

\begin{addmargin}[1em]{2em}
\emph{I think the great challenge of replicability is that we still don't run nearly enough replications\ldots We need to value replicability, and we need to create a niche for academics who work in replicability.%..I feel like there should be some journal, so some incentive that they can say that. Okay, if researchers are doing reproducible work, we are going to publish it for them.
} [Participant 5]
\end{addmargin}

\vspace{0.2cm}

\begin{addmargin}[1em]{2em}
\emph{If a journal would conditionally accept a certain project before data has been collected, just based on the pre-analysis plan, which includes a commitment to replicability.} [Participant 7]
\end{addmargin}

\subsubsection{Hybrid human-AI technologies to support reproducibility and replicability}

%We have addressed a very general query in this section because we want to explore how technology can support researchers in research reproducibility. We observed from the responses that one of the very helpful platforms is open science and that helps researchers a lot. Another very prominent platform is pre-registration which is appreciated by many researchers. 
We queried interview participants to understand their perspectives on opportunities for technologies, and in particular, human-AI technologies to support and enhance reproducibility and replicability. Participants were hopeful for technological interventions although generally preferred hybrid solutions. Seven of the eight participants expressed incomplete trust in AI-driven solutions alone.

\begin{addmargin}[1em]{2em}
\emph{I'm not sure that there is enough data on what is reproducible, and what isn't reproducible to train the AI model to do that correctly.} [Participant 2]
\end{addmargin}

\vspace{0.2cm}

\begin{addmargin}[1em]{2em}
\emph{I would expect the hybrid model would be more reliable\ldots just adding more predictors to a model tends to give you the better performance\ldots I expect it will add at least a little predictive power, and I'll be very keenly interested to see what the results of all those markets.} [Participant 5]
\end{addmargin}

\vspace{0.2cm}

%"Well, I'm a big fan of the open science framework sort of stuff that's going on. The availability of data, the availability of code.... So I think that maybe if we have computational reproducibility checks up front that would go a long way to increasing reproducibility"[Participant 5]

%"I think when I cite any paper, if I can see that score, I think it could be really helpful like It's more like I can trust that paper more"[Participant 6].

\begin{addmargin}[1em]{2em}
\emph{I believe in the markets\ldots Many people are convinced that markets are effective to aggregate information that the outcome is meaningful\ldots Probably my expectation would be the hybrid market performs most accurately.} [Participant 7].
\end{addmargin}

\subsubsection{Hybrid prediction market experience}

%This theme reflects participants' views of the hybrid market test run because they themselves have participated in this market. Based on their feedback we can make changes to our final hybrid market experiments. We have tried to analyze what are the things that may be less trustworthy for the participants or how we can make the experiments more user-friendly. We got very mixed reviews about this application which is very interesting to observe. Some participants thought the market was a bit longer and there was no volatility.
Finally, we collected participants' inputs on their experience during the hybrid markets and solicited suggestions for improvements to the platform and/or methodology moving forward. Most participants were enthusiastic and reported the experience as fun. Several highlighted details of the UI that could be improved. All participants felt the market duration was too long. We had selected a 2-hour window to afford best flexibility to participants logging in from around the world. We will revisit this in next steps.

%"I think confidence scoring approaches could be useful but I think there's a danger that there will be a backlash um to some extent, you know there's always the danger that they get it wrong....
\begin{addmargin}[1em]{2em}
\emph{To me, it seems that the market lasted quite a while\ldots It was really just slowly trending in one direction versus another direction\ldots It was really a lot of fun. Actually, I quite enjoyed it.} [Participant 1]
\end{addmargin}

%\vspace{0.2cm}

%\begin{addmargin}[1em]{2em}
%\emph{It is interactively interesting or understandable. I think it is a good interface, I think, better than I've seen for other experiments\ldots Another thing that I would have liked to do is instead of just buying and selling one share at a time, I would have liked to do it in bulk.} [Participant 2]
%\end{addmargin}

%\vspace{0.2cm}

%\begin{addmargin}[1em]{2em}I think that it is a long time for uh for the market, so you can. Maybe in the future. My suggestion just is maybe short on it to.} [Participant 4]
%\end{addmargin}

%\vspace{0.2cm}

%\begin{addmargin}[1em]{2em}It was actually quite nice. I like the interface as well....while in the AI market if I recall correctly, there were slide-upward or slide-down trends in these faces, and it was never a completely straight line...."[Participant 7]
%\end{addmargin}

\section{Conclusion}

We have described pilot studies with a prototype hybrid prediction market for replication prediction. This work in progress offers proof of viability of collaborative human-AI technology for the evaluation of published scientific claims. Although we pilot this approach in the context of replication prediction, we suggest that the hybrid market offers a new avenue for hybrid human-AI applicable to a broad set of tasks for which neither human- or machine-driven approaches alone are sufficient. Post-market interviews with participants highlight opportunities and challenges for this work. % of a larger project for developing and evaluating Human-AI collaboration for research reproducibility which is very unique. The 
%Our findings shed light on many aspects in which we can work further to improve our system. The purpose of this study is to understand how humans participating alongside autonomous agents impact outcomes of an artificial prediction market for evaluating the reproducibility of research claims. The proposed research could have an impact on society in broad ways because scientific research reproducibility is one of the most sensitive areas for researchers nowadays. This will create an APP that will give a confidence score for the published papers which can be used as a measurement for reproducibility this may help them during their literature review. A lot of money and time can be saved. The goal of the interview study was to gain insights about opinions, strategies, and viewpoints that work for the researchers to deal with hybrid markets and replication. Many interesting insights and findings were highlighted in our results which are really new in this domain.

\bibliographystyle{abbrv}
\bibliography{references}

\end{document}